\documentclass[12pt]{article}
\usepackage{mathrsfs}
\usepackage{amsmath,amssymb,array}
\usepackage{graphicx}
\usepackage{multicol}
\usepackage{indentfirst}
\numberwithin{equation}{section}

\begin{document}

\begin{titlepage}
\renewcommand{\thefootnote}{\fnsymbol{footnote}}

\begin{center}
{\LARGE \bf Generalized Thermodynamic Properties of  Lifshitz black holes}

\vspace{1.0cm}

{Keita Kaniba Mady$^{a,b}$}\footnote{E-mail: madyfalaye@gmail.com}, Dicko  Younouss Ham\`{e}ye$^{a, c, d,}$\footnote{E-mail: yunusshdicko@yahoo.fr}
\\

\vspace{.5cm}
{\it\small {$^{a}$ Centre de Calculs, de Mod\`{e}lisations et de Simulations, FST, Bamako,\\
$^{b}$ D\'{e}partement de Physique, Facult\'{e} des Sciences et Techniques, Universit\'{e} des Sciences, des Techniques et des Technologies de Bamako, Mali,}\\
$^{c}$ Institut de Consultation en Education,
\\$^{d}$ Effort pour le D\'{e}veloppement Int\'{e}gr\'{e} du Gourma. }\\
\vspace{.3cm}
\today
\end{center}\vspace{1.5cm}

\centerline{\textbf{Abstract}}\vspace{0.5cm}

The generalized effective potential for motion in Lifshitz geometry (with arbitrary value of both the dynamical exponent $z$ and $f(r)$) is presented. Based on this effective potential energy, we shall show that bounded orbits are forbidden for positive values  of $z$ and allowed  for negative values of  $z$. The stability of these bounded orbits are analysed in full detail. Besides, by comparison, we quote a theorem which should be very helpful in searching thermodynamical properties of these gravitational black holes. Based on this theorem, generalized thermodynamic properties of Lifshitz black holes are computed. Examples are given where the entropy still satisfies the area law. The calculated mass, temperature and entropy satisfy both the first thermodynamic principle of black holes and a Smarr formula.

\end{titlepage}
\setcounter{footnote}{0}


\section{Introduction}

In high energy physics, Maldacena \cite{M} conjectured that large $N$, strong 't Hooft \cite{tH} coupling, limit of gauge theory has a description in terms of supergavity solution. The solutions to this supergravity theory is used to approximately solve strongly conformal field theory in a $1/N$ expansion. Although this correspondence was established between Conformal field theory and AdS spacetime, many generalizations of this gauge/gravity duality to others branches of physics have been proposed in the literature. The investigations of some of these approaches yielded significant results in the recent years \cite{GKP, W, dTB, EKSS, DP1,KM, KKSS, Z, VS, KD, KD2, WF}.

Field theories describing condensed matter systems generally exhibit, at nearby their
critical points, an anisotropic (between spatial and temporal coordinates) scaling symme-
try : $t\rightarrow \lambda^{z}t$ and $x\rightarrow \lambda x$, where $z\neq1$ is called the dynamical critical exponent. In the vicinity of these critical points, these condensed matter systems are strongly coupled. The extension of AdS/CFT duality to such a Non-Relativistic field theories opened-up many more interesting avenues for investigating issues, which are untractable otherwise, of physical systems in the real world. From this inspection, two spacetimes asymptotically emerged as background for gravity dual to non-relativistic conformal field theories: Lifshitz spacetimes and Schrodinger spacetimes \cite{kous, sha}. The black holes which asymptote Lifshitz spacetimes are the so-called Lifshitz black holes \cite{kous2, rong, yl, ayn, daw, gim, hai, bra, ber, do, hma}.

The thermodynamic properties of these Black holes as well as particles motion in their vicinity are still an unfinished business which  call for an appropriate settlement. For instance, the definition  of the mass and energy (conserved charged) of these black holes still need to be clarified see e.g. (\cite{eloy1, yun, wgb, zan, hyun} and references therein). What is unusual, when the parameters of the black hole are chosen appropriately, we can have the entropy and the mass both negative or zero entropy while non-zero temperature \cite{rong}. As far as motion of particles in the vicinity of these black holes is concerned, we are very far from writing the final verbatim report down. Some aspects of this last problem dealing with very specific kind of $f(r)$ or $z$ have been reported in the literature \cite{ber, hus, marco}. To our knowledge, no analyze for motion of particles has been done in these black holes with full generality. In addition, it has even been shown that bounded orbits are  forbidden in this background under specific conditions \cite{marco}. What is more, the radial photon cannot even determined the presence of the black hole in certain region \cite{vill}. It is therefore necessary to further our understanding of these gravitational background. This manuscript intends to push forward our comprehension about these black holes.

The remaining of this article is organized as follow: In the second section, we shall utilize Hamilton-Jacobi equation to find the generalized effective potential for massive as well as massless particles in these gravitational field and meantime quote a Theorem which should be helpful in future. We will then go on (the third section) by analyzing in full detail the behavior of this generalized effective potential for specific form of $f(r)$ and $z$. In the fourth section, we shall show the utility of our theorem, by computing the generalized thermodynamic properties (Mass, entropy, heat capacity, Temperature) of Lifshitz black holes. Finally, we conclude our discussion in the last section\footnote{We set $G=c=1$ throughout this manuscript.}.

\section{Hamilton-Jacobi Equation}
\subsection{Massive particles}
The line element of Lifshitz black holes backgrounds in $3+1$-dimensions is given by
\begin{equation}\label{metric}
    ds^{2}=-r^{2z}f(r)dt^{2}+\frac{1}{r^{2}f(r)}dr^{2}+r^{2}d\theta^{2}+r^{2}sin^{2}\theta d\phi^{2}
\end{equation}
where the function $f(r)$ goes to $1$ as $r\rightarrow \infty$. The highest root of the equation $f(r)=0$ is known as  the radius $r_{H}$ of the horizon.

Several forms of this spherically symmetric metric have been found to be solution to Einstein's equations for generic quadratic curvature gravity theory. The boundary of these black holes is located at infinite $r$, and the metric reduces to the ones of Lifshitz spacetimes, which read

\begin{equation}\label{metric2}
    ds^{2}=-r^{2z}dt^{2}+\frac{1}{r^{2}}dr^{2}+r^{2}d\theta^{2}+r^{2}sin^{2}\theta d\phi^{2}
\end{equation}

The properties (such as singularity and symmetry) of this metric are well-known concepts \cite{kous, sha}.

As a general rule, the motion of particles in centrally symmetric field always occurs in a single "plane" passing through the origin; now defining this plane to be the equatorial plane characterized by $\theta=\frac{\pi}{2}$. The trajectory, for a test particle of rest mass $m$ traveling in this reduced Lifshitz curved space, is given by the equation of Hamilton-Jacobi :
 \begin{equation}\label{hj}
    g^{ij}\frac{\partial S}{\partial x^{i}}\frac{\partial S}{\partial x^{j}}+m^{2}=0
 \end{equation}

 Using the $g^{ij}$ derived from the reduced Lifshitz black hole metric, eq.(\ref{hj}) becomes
 \begin{equation}\label{hj2}
   \frac{1}{r^{2z}f(r)}\left(\frac{\partial S}{\partial t}\right)^{2}-r^{2}f(r)\left(\frac{\partial S}{\partial r}\right)^{2}-\frac{1}{r^{2}}\left(\frac{\partial S}{\partial \phi}\right)^{2}-m^{2}=0
 \end{equation}

  To solve Eq.(\ref{hj2}), we shall apply the 'method of separation of variables' by writing $S$ in the form:
 \begin{equation}\label{s}
    S=-\emph{E}_{0}t+M\phi+S_{r}(r),
 \end{equation}

where $\emph{E}_{0}$ and $M$ are the constants of energy and angular momentum,  respectively. Plugging eq.(\ref{s}) back into eq.(\ref{hj2}) and solving it in term of $S_{r}(r)$, one finds:
\begin{equation}\label{sr}
    S_{r}(r)= \int \left(\frac{1}{r^{z+1}f(r)}\sqrt{\emph{E}^{2}_{0}-r^{2z}f(r)m^{2}\left(1+\frac{M^{2}}{m^{2}r^{2}}\right)}\right).dr
\end{equation}

The dependence $r=r(t)$ is obtained by requiring that $\frac{\partial S}{\partial \emph{E}_{0}}=Const$, from which one gets , in differential form (by differentiating both sides of this equality),
\begin{equation}\label{diff}
    d\left(\frac{\partial S}{\partial \emph{E}_{0}}\right)=0\Longleftrightarrow dt=\frac{\emph{E}_{0}}{r^{z+1}f(r)}\frac{dr}{\sqrt{\emph{E}^{2}_{0}-r^{2z}f(r)m^{2}\left(1+\frac{M^{2}}{m^{2}r^{2}}\right)}}
\end{equation}

Eq.(\ref{diff}) is usually written in the form:
\begin{equation}\label{eff}
    \frac{\emph{E}_{0}}{r^{z+1}f(r)}\frac{dr}{dt}=\frac{1}{\emph{E}_{0}}\sqrt{\emph{E}^{2}_{0}-r^{2z}f(r)m^{2}\left(1+\frac{M^{2}}{m^{2}r^{2}}\right)}=\frac{1}{\emph{E}_{0}}\sqrt{\emph{E}^{2}_{0}-V_{eff}^{2}(r)}
\end{equation}
The function
\begin{equation}\label{epo}
V_{eff}(r)=m r^{z}\sqrt{f(r)\left(1+\frac{M^{2}}{m^{2}r^{2}}\right)}
\end{equation}
is what we call the generalized  effective potential for motion of massive particle in Lifshitz black holes\footnote{Some authors take $V^{2}_{eff}(r)$ as the effective potential\cite{bd, marco}; but in this manuscript, we will closely follow Landau et al.\cite{landau} and Misner et al.\cite{misner}  }.

The trajectory of the particles is determined by the equation $\frac{\partial S}{\partial M}=Const$, so that (by differentiating this equality)
\begin{equation}\label{tra}
    \phi=\int \frac{r^{z-3}M}{\sqrt{\emph{E}^{2}_{0}-r^{2z}f(r)m^{2}\left(1+\frac{M^{2}}{m^{2}r^{2}}\right)}}dr
\end{equation}

\subsection{Massless Particles}
In regard to motion of massless particles, the Hamilton-Jacobi equation eq.(\ref{hj}), after the following substitutions
\begin{itemize}
  \item $S\rightarrow \psi$
  \item $m\rightarrow 0$
  \item $\emph{E}_{0}\rightarrow \omega$
  \item $\varrho\rightarrow \frac{M}{\omega}$
\end{itemize}

 becomes the so-called eikonal equation. We solve it in the form
 \begin{equation}\label{ei}
    \psi=-\omega t+ \omega \varrho \varphi+ \psi_{r}(r)
 \end{equation}
 where
 \begin{equation}\label{eik}
    \psi_{r}(r)=\int \left(\frac{\omega}{r^{z+1}}\sqrt{\frac{1}{f(r)^{2}}-r^{2(z-1)}\frac{\varrho^{2}}{f(r)}}\right).dr
 \end{equation}
 Additionally, the path of the ray (massless particles) resulting from eq.(\ref{tra}) is given by
 \begin{equation}\label{rtra}
    \phi=\int \frac{r^{z-3}\varrho}{\sqrt{1-r^{2(z-1)}f(r)\varrho^{2}}}dr
 \end{equation}
\subsection{Theorem}
Thermodynamic properties of Lifshitz black holes should reduced to their symmetric spherical black holes counterpart when the following limits are taken:
\begin{itemize}
  \item $z\rightarrow 1$
  \item $f(r)\rightarrow \frac{h(r)}{r^{2}}$
\end{itemize}
\subsubsection{Proof} The proof of this theorem is trivial, since the metric in these limits is simply the one of spherical symmetric black hole spacetimes.
\begin{equation}\label{metricsy}
    ds^{2}=-h(r)dt^{2}+\frac{1}{h(r)}dr^{2}+r^{2}d\theta^{2}+r^{2}sin^{2}\theta d\phi^{2}
\end{equation}
Some may object that this form of $f(r)$ is in conflict with the constraint that the function $f(r)$ goes to $1$ as $r\rightarrow \infty$. Though this is true, the theorem has to be taken as a guidance principle when evaluating thermodynamics properties of these black holes.

So far all the thermodynamic properties of these black holes (defined without problem) satisfy this theorem. For example, the temperature $T_{Lbh}=\frac{r_{+}^{z+1}f'(r_{+})}{4\pi}$, the generalized effective potential eq. (\ref{epo}), as well as the trajectories eqs. (\ref{tra}) and (\ref{rtra}) reduced to theirs symmetric spherical counterparts.
\section{Bounded Orbits}
\subsection{Positive Values of $z$}
We begin our study of bounded orbits by considering the case of positive values of $z$. Namely, $z=2$, $z=4$ and the form $f(r)=1-\frac{r_{H}}{r}=1-\frac{1}{x}$. Bounded orbits are determined by the plot of the effective potential by using the concepts of turning points. Turning points are those special values of $r_{n} (n=1, 2, 3, ...)$ for which $E=V_{eff}(r_{n})$, i.e., $\frac{dr}{dt}=0$ at these points. If two non-vanishing turning points $r_{2}<r_{1}<\infty$ exist, the motion is said to be bounded in the interval $r_{2}<r<r_{1}$, otherwise the motion is unbounded\footnote{ See page 109 of "An Introduction to Lagrangian Mechanics, Alain J, Brizard"}. In term of the effective potential, we shall have $V_{eff}(r_{2})=V_{eff}(r_{1})$ for bounded orbits. For the chosen parameters, it is visible from the plot (Fig1) of the effective potential energy that this condition cannot be satisfied at all. This failure can be generalized to all $z>0$. That is, there is no bounded orbits for $z$ positive.
\begin{figure}[h]
\begin{center}
\includegraphics[width=10cm,clip=true,keepaspectratio=true]{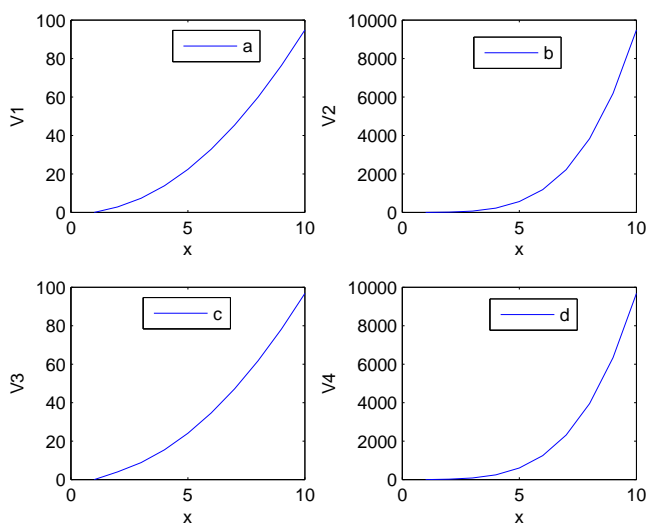}
\caption{\small The plot of $V_{eff}(r)$, where $f(r)=1-\frac{1}{x}$; $x=\frac{r}{r_{H}}$ and: $(-a)$$\sim$$(z, M, V1)=(2, 0, \frac{V_{eff}}{m r^{2}_{H}})$ ; $(-b)$$\sim$$(z, M, V2)=(4, 0, \frac{V_{eff}}{m r^{4}_{H}})$; $(-c)$$\sim$$(z, M, V3)=(2, 2 m r_{H}, \frac{V_{eff}}{m r^{2}_{H}})$;$(-d)$$\sim$$(z, M, V4)=(4, 2 m r_{H}, \frac{V_{eff}}{m r^{4}_{H}})$ .}
\end{center}\label{fig}
\end{figure}

The value $z=4$ is very special \cite{fw}.
\subsection{Negative value of $z$}
Asymptotically Lifshitz black holes have been found for $z<0$ \cite{ayn}. The effective potential is still the same, though these black holes are defined only in higher dimensions $D\geq 5$. Since one can always perform its computation in the equatorial plane. The plot of the effective potential for the same parameters as above shows that these values allow bounded orbits. From the form of the effective potential, it can be generalized that bounded orbits are allowed  for all negative values of $z$.
\begin{figure}[h]
\begin{center}
\includegraphics[width=10cm,clip=true,keepaspectratio=true]{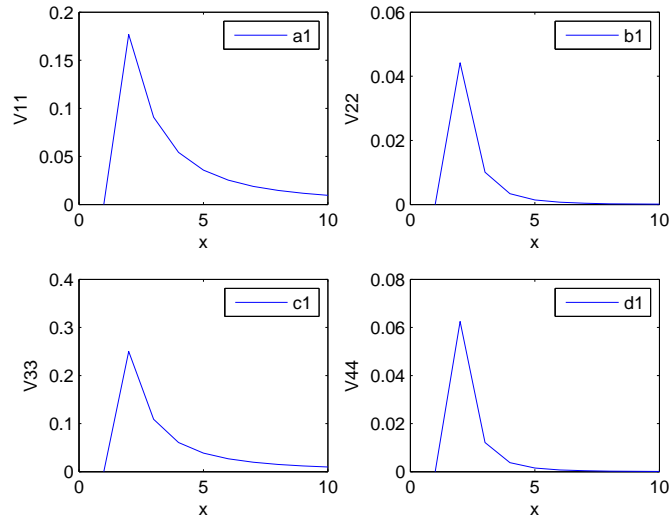}
\caption{\small The plot of $V_{eff}(r)$, where $f(r)=1-\frac{1}{x}$; $x=\frac{r}{r_{H}}$ and: $(-a1)$$\sim$$(z, M, V11)=(-2, 0, \frac{V_{eff}}{m}r^{2}_{H})$ ; $(-b1)$$\sim$$(z, M, V22)=(4, 0, \frac{V_{eff}}{m}r^{4}_{H})$; $(-c1)$$\sim$$(z, M, V33)=(2, 2 m r_{H}, \frac{V_{eff}}{m}r^{2}_{H})$;$(-d1)$$\sim$$(z, M, V44)=(4, 2 m r_{H}, \frac{V_{eff}}{m }r^{4}_{H})$ .}
\end{center}\label{fig}
\end{figure}

This fact is clearly demonstrated in  (fig2)
\subsubsection{Stability of these bounded orbits}
To study the stability, we shall find the derivative of the effective potential. The solutions of the equation $V^{\prime}_{eff}(r)=0$ determine the minima or the maxima of its plot . By using (\ref{epo})and $z=-\eta$ with $\eta>0$, this equation reduces to the following:
\begin{equation}\label{maxi}
    -2f(r)\left((\eta+1)M^{2}+\eta m^{2}r^{2}\right)+r\left(M^{2}+ m^{2}r^{2}\right)f^{\prime}(r)=0
\end{equation}
which in its turn reduced to its symmetric spherical counterpart in the conditions of the theorem:
\begin{equation}\label{maxisy}
   -2h(r)M^{2}+r\left(M^{2}+ m^{2}r^{2}\right)h^{\prime}(r)=0
\end{equation}
The solutions of eq.(\ref{maxi}), for a specific form of $f(r)$, determine the maxima or the minima of the effective potential.
\section{Mass and Entropy of Lifshitz Black holes}
\subsection{Computation}
A good place to start our discussion is to use the definition of conventional black holes mass. For the line element of conventional black holes backgrounds in $3+1$-dimensions of the form
\begin{equation}\label{conv}
    ds^{2}=-h(r)dt^{2}+\frac{1}{h(r)}dr^{2}+r^{2}d\theta^{2}+r^{2}sin^{2}\theta d\phi^{2},
\end{equation}
the mass of the black holes (thermodynamic mass)$M_{bh}$ is given by the formula
\begin{equation}\label{masscon}
    M_{bh}=\frac{1}{2}r^{2}_{+}h^{\prime}(r_{+}).
\end{equation}
Where $^{\prime}$ denotes derivative with respect to $r$ and $r_{+}$ is the horizon of the black holes.

Based on our theorem and the formula (\ref{masscon}), we propose the following formula for the mass of Lifshitz black holes
\begin{equation}\label{masslif}
    M_{lifbh}=\frac{1}{2}r^{z+3}_{+}f^{\prime}(r_{+}).
\end{equation}
Demanding that the Lifshitz black holes satisfy the first law of black holes thermodynamic
\begin{equation}\label{firstlaw}
    dM=TdS,
\end{equation}
one finds that the generalized entropy of Lifshitz black holes is given by
\begin{equation}\label{entropy}
    S_{Lifbh}=\pi (z+3)r^{2}_{+}+2\pi \left(r^{2}_{+}\ln(f'(r_{+}))\right)-4\pi \left(r_{+}h(r_{+})\right)+4\pi\int h(r_{+})dr_{+}.
\end{equation}
Where $h'(r_{+})=\ln (f'(r_{+}))$; we used eq.(\ref{masslif}) and the value of the temperature of Lifshitz black holes given by
\begin{equation}\label{templif}
T_{Lifbh}=\frac{r_{+}^{z+1}f'(r_{+})}{4\pi}.
\end{equation}
Eqs.(\ref{masslif}), (\ref{entropy})and (\ref{templif}) are what we called the generalized Mass, Entropy and Temperature of Lifshitz black holes, respectively.
\subsection{Stability}
To study the stability of Lifshitz black holes in our setting, one simmply calculates the heat capacity $C$ which is given by
\begin{equation}\label{heat}
    C=\frac{\partial M}{\partial T}=\frac{\partial M}{\partial r_{+}}\frac{\partial r_{+}}{\partial T}=\frac{\partial M}{\partial r_{+}}\left(\frac{\partial T}{\partial r_{+}}\right)^{-1}
\end{equation}
This give us the generalized heat capacity of Lifshitz black holes as
\begin{equation}\label{heatg}
    C=2\pi r^{2}_{+}\frac{(z+3)f'(r_{+})+r_{+}f''(r_{+})}{(z+1)f'(r_{+})+r_{+}f''(r_{+})}
\end{equation}

\subsection{Smarr formula}
Let us exemplify our computation by using very specific form of $f(r)$, but with unspecified value of $z$. We start with the most popular form $f(r)=1-\frac{r_{+}}{r}$. One easily computes the above parameters as
\begin{equation}\label{fex}
    T_{Lifbh}=\frac{r^{z}_{+}}{4\pi}\qquad\text{,}\qquad M_{Lifbh}=\frac{1}{2}r^{z+2}_{+}\qquad\text{and}\qquad S_{Lifbh}=\pi (z+1)r^{2}_{+}
\end{equation}
and the heat capacity for this black hole is given by
\begin{equation}\label{heat1}
     C=2\pi r^{2}_{+}\frac{z+1}{z-1}.
\end{equation}

These results give us a very beautiful Smarr formula of the form
\begin{equation}\label{smarr1}
    M=\frac{2}{z+1}T.S.
\end{equation}
This a confirmation of the anisotropic version of the Smarr formula \cite{eloy1}.

For this black hole, it is clear that the entropy is positive definite when $z+1>0$, under this condition, the heat capacity is positive-definite for $z>1$ ; therefore, this black hole is stable when $z>1$.

Our second example dealt with the form $f(r)=1-\frac{r^{\eta}_{+}}{r^{\eta}}$. This is another popular form with has been also used in the literature.
For this one, the parameters of the black holes are
\begin{equation}\label{sex}
    T_{Lifbh}=\eta\frac{r^{z}_{+}}{4\pi}\qquad\text{,}\qquad M_{Lifbh}=\frac{\eta}{2} r^{z+2}_{+}\qquad \text{and}\qquad S_{Lifbh}=\pi (z+2-\eta)r^{2}_{+}
\end{equation}
and it has a heat capacity given by
\begin{equation}\label{heat2}
   C=2\pi r^{2}_{+}\frac{z+2-\eta}{z-\eta}
\end{equation}

These last results give us another very beautiful Smarr formula of the form
\begin{equation}\label{smarr2}
    M=\frac{2}{z+2-\eta}T.S.
\end{equation}
Here, there is a violation the anisotropic property of the Smarr formula.

For this last black hole, it should be stressed that the entropy is positive definite when $z+2>\eta$, and due to this circumstance the heat capacity is positive when $z>\eta$; therefore, this black hole is also stable $z>\eta$.

\section{Conclusions and perspectives}
The aims of this  manuscript are twofold: The first part is devoted to the motion of particles in analytic Lifshitz black holes. In this respect, our article clearly demonstrated that bounded orbits are allowed in these black holes; One can use our formula eq.(\ref{maxi}) to localize the maxima or the minima of these bounded orbits for any specific form of $f(r)$. In addition, the eqs. (\ref{tra}) and (\ref{rtra}) can be used to determine the shift of the perihelion of the orbits in these black holes. In the second part, we have unquestionably shown that our theorem is a very powerful asset when evaluating the correct form of the thermodynamic properties of these black holes. This can be seen from the generalize results of the different parameters we get for  these black holes.
\subsection*{Acknowledgments}
We thank all the members of CCMS for enlightening discussions and remarks. Keita would like to thank Prof.Wu Feng for his tireless encouragement and help; he is deeply indebted to Prof. Yi Ling and Dr. Wu Jian-Ping for all their help.


\end{document}